\documentclass{aa}

\usepackage{graphicx}
\usepackage{txfonts}
\def\arcsec{\ifmmode {''} \else ${''}$\fi}
\def\arcmin{\ifmmode {'} \else ${'}$\fi}
\def\deg{\ifmmode {^\circ} \else ${^\circ}$\fi}
\def\cc{\ifmmode {\rm cm}^{-3} \else cm$^{-3}$\fi}
\def\cl{\ifmmode {\rm cm}^{-2} \else cm$^{-2}$\fi}
\def\pcm2{\ifmmode {\rm cm}^{-2} \else cm$^{-2}$\fi}
\def\micron{\ifmmode \mu{\rm m} \else $\mu$m\fi}
\def\kms{\ifmmode {\rm km\,s}^{-1} \else km\,s$^{-1}$\fi}
\def\kmps{\ifmmode {\rm km\,s}^{-1} \else km\,s$^{-1}$\fi}
\def\Hubble{\ifmmode {\rm km\,s}^{-1}\,{\rm Mpc}^{-1}
        \else km\,s$^{-1}$\,Mpc$^{-1}$\fi}
\def\ergsec{\ifmmode {\rm ergs\;s}^{-1} \else ergs s$^{-1}$\fi}
\def\ergcms{\ifmmode {\rm ergs\,cm}^{-2}\,{\rm s}^{-1}
          \else ergs\,cm$^{-2}$\,s$^{-1}$\fi}
\def\ergcmsA{\ifmmode {\rm ergs\,cm}^{-2}\,{\rm s}^{-1}\,{\rm \AA}^{-1}
          \else ergs\,cm$^{-2}$\,s$^{-1}$\,\AA$^{-1}$\fi}
\def\ergcmsHz{\ifmmode {\rm ergs\,cm}^{-2}\,{\rm s}^{-2}\,{\rm Hz}^{-1}
          \else ergs\,cm$^{-2}$\,s$^{-1}$\,Hz$^{-1}$\fi}
\def\Msun{\ifmmode M_{\odot} \else $M_{\odot}$\fi}
\def\Lsun{\ifmmode L_{\odot} \else $L_{\odot}$\fi}
\def\qo{\ifmmode q_{0} \else $q_{0}$\fi}
\def\Ho{\ifmmode H_{0} \else $H_{0}$\fi}
\def\ltsim{\raisebox{-.5ex}{$\;\stackrel{<}{\sim}\;$}}
\def\gtsim{\raisebox{-.5ex}{$\;\stackrel{>}{\sim}\;$}}
\begin{document}

\title{Multiwavelength campaign on Mrk 509}
\subtitle{XVI. Continued HST/COS monitoring of the far-ultraviolet spectrum}

\author{ G. A. Kriss\inst{\ref{inst1}}
	\and
          N. Arav\inst{\ref{inst2}}
          \and
	  D. Edmonds\inst{\ref{inst3}}
          \and
	  J. Ely\inst{\ref{inst1}}
          \and
          J.S. Kaastra\inst{\ref{inst4},\ref{inst5}}
          \and
          S. Bianchi\inst{\ref{inst9}}
          \and
          M. Cappi\inst{\ref{inst10}}
          \and
          E. Costantini\inst{\ref{inst4}}
           \and
           J. Ebrero \inst{\ref{inst12}}
          \and
          M. Mehdipour\inst{\ref{inst4}}
	  \and
          S. Paltani\inst{\ref{inst13}}
          \and
          P. Petrucci\inst{\ref{inst14}}
	  \and
          G. Ponti\inst{\ref{inst11}}
}

\institute{
	Space Telescope Science Institute,
        3700 San Martin Drive, Baltimore, MD, 21218, USA \email{gak@stsci.edu}\label{inst1}
\and
    Department of Physics, Virginia Tech, Blacksburg, VA 24061, USA\label{inst2}
\and
	The Pennsylvania State University, Department of Physics,
	76 University Drive,
        Hazleton, PA, 18202, USA\label{inst3}
\and
    SRON Netherlands Institute for Space Research,
    Sorbonnelaan 2, 3584 CA Utrecht, The Netherlands\label{inst4}
    \and
    Leiden Observatory, Leiden University, PO Box 9513, 2300 RA Leiden, the Netherlands\label{inst5}
    \and
    Dipartimento di Matematica e Fisica, Universit\`a degli Studi Roma Tre, via della
Vasca Navale 84, I-00146 Roma, Italy\label{inst9}
    \and
    INAF-IASF Bologna, Via Gobetti 101, I-40129 Bologna, Italy\label{inst10}
    \and
    European Space Astronomy Centre, P.O. Box 78, E-28691 Villanueva de la Ca\~{n}ada, Madrid, Spain\label{inst12}
    \and
    Department of Astronomy, University of Geneva, 16 Ch. d'Ecogia, 1290 Versoix, Switzerland\label{inst13}
    \and
    Univ. Grenoble Alpes, CNRS, IPAG, 38000 Grenoble, France\label{inst14}
    \and
    INAF-Osservatorio Astronomico di Brera, Via E. Bianchi 46, I-23807 Merate (LC), Italy\label{inst11}
}

\date{
	Accepted for publication, January 22, 2019
}


  \abstract
  {}
   {To elucidate the location, physical conditions, mass outflow rate, and
   kinetic luminosity of the outflow from the active nucleus of the Seyfert 1
   galaxy Mrk 509
   we used coordinated ultraviolet and X-ray spectral observations
    in 2012 to followup our lengthier campaign conducted in 2009.
   }
   {We observed Mrk 509 with the Cosmic Origins Spectrograph (COS) on the
    {\it Hubble Space Telescope (HST)} on 2012-09-03 and 2012-10-11
    coordinated with X-ray observations using the
    High Energy Transmission Grating on the {\it Chandra X-ray Observatory}.
    Our far-ultraviolet spectra used grating G140L on
    COS to cover wavelengths from 920--2000 \AA\ at a resolving power of
    $\sim2000$, and gratings G130M and G160M to cover 1160--1750 \AA\ at a
    resolving power of $\sim15,000$.}
   {We detect variability in the blue-shifted UV absorption lines
    on timescales spanning 3--12 years. The inferred densities in the
    absorbing gas are greater than log $n \rm~cm^{-3} \sim 3$.
     For ionization parameters ranging over log $U = -1.5 \rm~to~-0.2$, we
    constrain the distances of the absorbers to be closer than 220 pc to the
    active nucleus.
    }
   {The impact on the host galaxy appears to be confined to the nuclear region.
   }


\authorrunning{G. A. Kriss et al.}

\titlerunning{HST/COS observations of Mrk 509}

\maketitle
%

\section{Introduction}

As one of the brightest Seyfert galaxies ($m_V = 13.5$ \cite{McAlary83};
$z = 0.034397$ \cite{Fisher95}),
Mrk 509 has been the target of
many campaigns to understand its physical characteristics in greater
detail as an aid to understanding the physics of active galactic nuclei (AGN)
in general.
With a bolometric luminosity of
$\rm M_{bol} = 2.3 \times 10^{45}~\rm erg~s^{-1}$ \citep{Runnoe12},
it lies on the Seyfert/quasar classification boundary \citep{Kopylov74}.
Campaigns with the {\it International Ultraviolet Explorer (IUE)}
established its ultraviolet variability \citep{Chapman85}.
Ground-based reverberation mapping using optical spectra
\citep{Peterson98, Kaspi00, Peterson04}
established a black-hole mass of $1.1 \times 10^8~\Msun$ \citep{Bentz15}.
The earliest IUE observations \citep{Wu80, York84}
revealed the blue-shifted ultraviolet absorption characteristic of AGN outflows.
These features were
studied at more detail and at higher spectral resolution with the
{\it Far Ultraviolet Spectroscopic Explorer} (FUSE) \citep{Kriss00}
and the Space Telescope Imaging Spectrograph (STIS) on the
{\it Hubble Space Telescope (HST)} \citep{Kraemer03}.
Blue-shifted absorption indicative of outflowing gas also appears in
X-ray spectra \citep{Yaqoob03}.

\begin{table*}
  \centering
	\caption[]{COS Observations of Mrk~509}
	\label{ObsTbl}
\begin{tabular}{l c c c c c}
\hline\hline       
Data Set Name & Grating/Tilt  & Date & Start Time & Start Time & Exposure Time\\
              &               &      &    (GMT)   &  (MJD)     &       (s)\\
\hline
lc0t01010 & G130M/1309 & 2012-09-03 & 10:01:30 & 56173.417713 & 1642 \\
lc0t01020 & G160M/1577 & 2012-09-03 & 10:36:09 & 56173.441775 & 2707 \\
lc0t01030 & G140L/1280 & 2012-09-03 & 13:00:12 & 56173.541810 & 5041 \\
lc0t02010 & G130M/1309 & 2012-10-11 & 12:25:24 & 56211.517643 & 1642 \\
lc0t02020 & G160M/1577 & 2012-10-11 & 13:00:03 & 56211.541706 & 2707 \\
lc0t02030 & G140L/1280 & 2012-10-11 & 15:24:09 & 56211.641775 & 5041 \\
\hline                  
\end{tabular}
\end{table*}

The galaxy-wide nature of this outflow is manifested by extended, blue-shifted
emission from [\ion{O}{iii}] in optical long-slit spectra \citep{Phillips83},
HST images \citep{Fischer15}, and ground-based integral-field unit (IFU)
observations \citep{Liu15}.
Such outflows powered by the central AGN are often invoked in models of
galaxy formation in order to ameliorate several related issues.
Feedback instigated by an outflow may suppress star formation or even expel
gas from the host galaxy
\citep{Silk98,King03,Ostriker10,Soker10,Faucher12,Zubovas14,Thompson15}.
This in turn can link the properties of the host galaxy to those of its central
black hole leading to the correlation between the black-hole mass and the
central velocity dispersion of the galaxy bulge \citep{DiMatteo05,Hopkins10}.
These models require the outflow to tap into
0.5--5\% of the luminosity radiated by the AGN.
Measuring the mass flux and the kinetic luminosity of the outflow is therefore
central to testing such models of feedback.
This, in turn, requires knowledge of the location and physical conditions in
the outflow.

In 2009 we undertook an extensive monitoring
campaign on Mrk 509 using {\it XMM-Newton}, {\it Chandra},
{\it HST}, {\it Swift}, and {\it INTEGRAL} \citep{Kaastra11a}.
This campaign established the location of several of the components of the
UV outflow \citep{Kriss11b, Arav12} and the X-ray absorbers \citep{Kaastra12}.
The UV absorbers are characterized by seven discrete velocity components
ranging from $-700$ to $+200~\rm km~s^{-1}$ relative to the systemic
velocity of the host galaxy. \cite{Arav12} set lower limits of 100--200 pc
for the location of these absorbers.
The X-ray absorbers lie in the same velocity range, and they comprise
six distinct components in velocity and ionization state
with total column densities of $\rm 0.8-6.3 \times 10^{20}~cm^{-2}$
\citep{Detmers11}.
\cite{Kaastra12} showed that the X-ray absorbers lie at distances ranging from
5 pc to 3 kpc.

In this paper we describe a longer-timescale continuation
of our original monitoring program.
In the fall of 2012 we obtained additional COS spectra of Mrk 509,
including far-UV optimized grating tilts that yield spectra down to the
galactic Lyman limit. These spectra overlap the wavelength range of the
original FUSE spectra \citep{Kriss00, Kriss11b}.
These observations were coordinated with {\it Chandra} HETGS
spectra, previously discussed by \cite{Kaastra14b}.
In this paper we describe our COS observations and our data reduction methods
in \S2.
\S3 compares these new spectra to those from prior campaigns.
\S4 discusses the implications of our new observations, focusing in particular
on more distant absorbing gas that exhibits longer-term variability.
\S5 summarizes our results.

\section{Observations and Data Reduction}

Our {\it HST}/COS spectra of Mrk 509 were obtained on 2012-09-03 and
2012-10-11, close in time to the coordinated {\it Chandra} HETGS
observations of 2012-09-04 and 2012-09-09 described by \cite{Kaastra14b}.
\cite{Green12} describe COS and its in-orbit performance.
We used two orbits for each COS visit using gratings G130M, G160M, and G140L.
With the medium resolution gratings (resolving power $\sim 15,000$), we covered
the 1160--1750 \AA\ wavelength range.
The low resolution grating G140L covered a broader range in wavelength,
920--2100 \AA, at lower resolution, with R$\sim$3,000.
With each grating we used multiple focal-plane settings (FP-POS) to place the
spectrum on independent locations of the detector in order to mitigate
against spectral features introduced by detector artifacts and grid-wire
shadows. All observations are summarized in Table \ref{ObsTbl}.

We obtained the data from the archive and processed it using v2.18.5 of the
COS calibration pipeline supplemented by custom flat-field corrections and
wavelength calibrations as described by \cite{Kriss11b}, but specifically
tailored to observations at COS Lifetime Position 2 (LP2).
We also made zero-point corrections (which were $< 1$ pixel) determined by
cross-correlating each exposure with the prior HST/STIS spectra of Mrk~509
before merging them into a final merged spectrum from the two visits.
During the 2012-09-03 observation Mrk 509 was slightly brighter than the
2009 observation of \cite{Kriss11b} with a flux of F(1367 \AA)=
$9.37 \times 10^{-14}~\rm erg~cm^{-2}~s^{-1}~\AA^{-1}$ compared to
$8.83 \times 10^{-14}~\rm erg~cm^{-2}~s^{-1}~\AA^{-1}$.
On 2012-10-14 the flux was 20\% fainter, with
F(1367 \AA)= $7.67 \times 10^{-14}~\rm erg~cm^{-2}~s^{-1}~\AA^{-1}$.
All three spectra look nearly identical, and so we do not show them here.

Despite the similarity of the 2012 COS spectra and those obtained in 2009,
examining changes in the intrinsic absorption line troughs requires careful
attention to the instrumental properties, particularly the spectral resolution.
Since the two observations were obtained at two different lifetime positions on
the COS UV detector, they differ in resolution. Thus we can not directly
compare the actual calibrated spectra.
Therefore, as described by \cite{Kriss11b}, we deconvolved the COS 2012
spectra using lines-spread functions (LSFs) appropriate to LP2
\citep{Duval13}.
Each separate exposure of each COS spectral region was deconvolved
using an LSF appropriate for the grating tilt for that exposure and
as specified for the central wavelengths of the Ly$\alpha$, \ion{N}{v}, and
\ion{C}{iv} emission lines.
These deconvolved spectra can be compared on an equal footing.
For both sets of observations, the deconvolution ameliorates the broad, shallow
wings of the LSF, rendering narrow absorption features deeper, and making the
central troughs of black, saturated absorption features
consistent with zero flux.

\begin{figure*}[!tbp]
  \centering
   \includegraphics[width=17cm, angle=0, scale=1.0, trim=0 485 0 0]{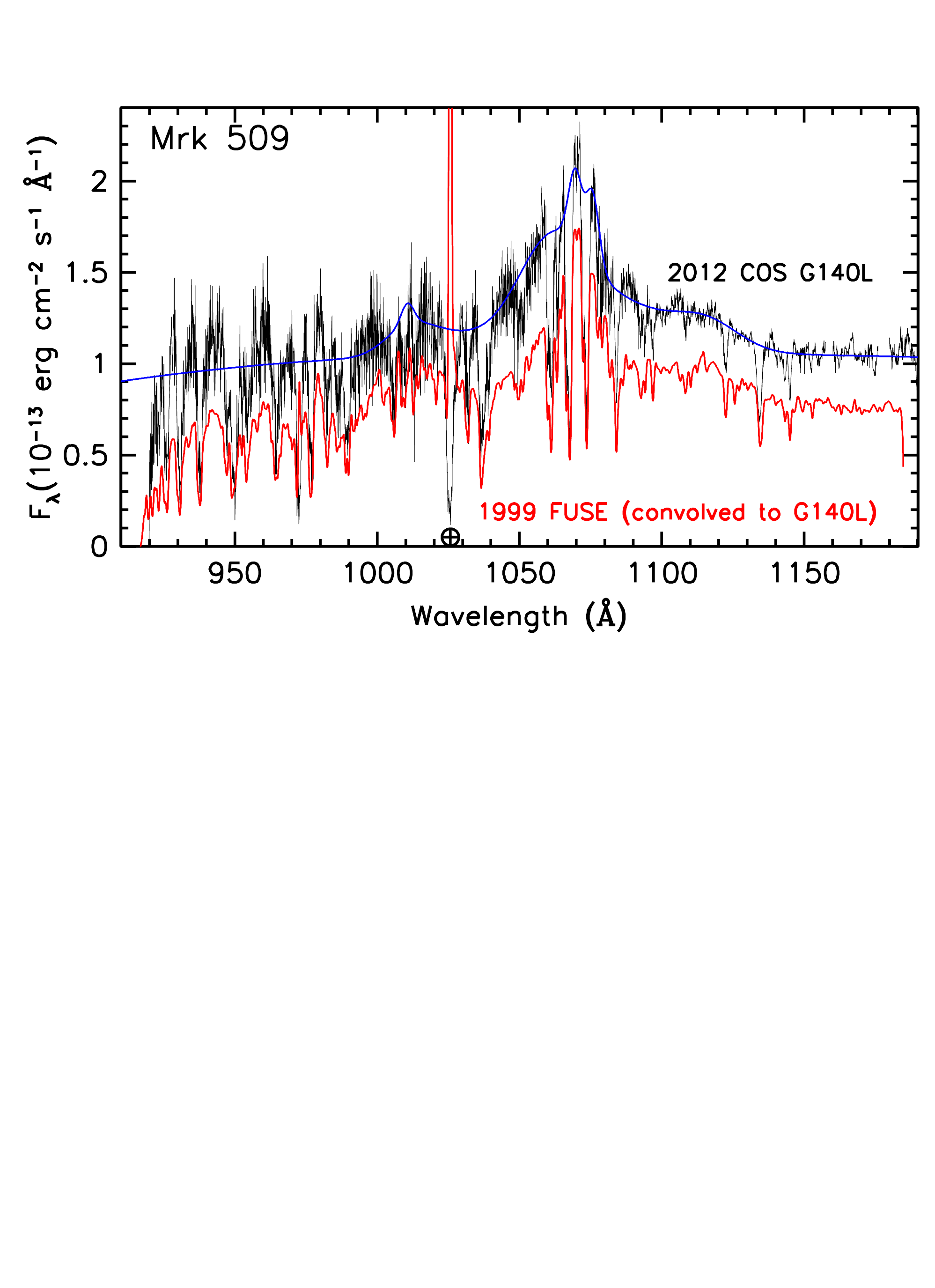}
\vskip 20pt
   \includegraphics[width=17cm, angle=0, scale=1.0, trim=0 485 0 0]{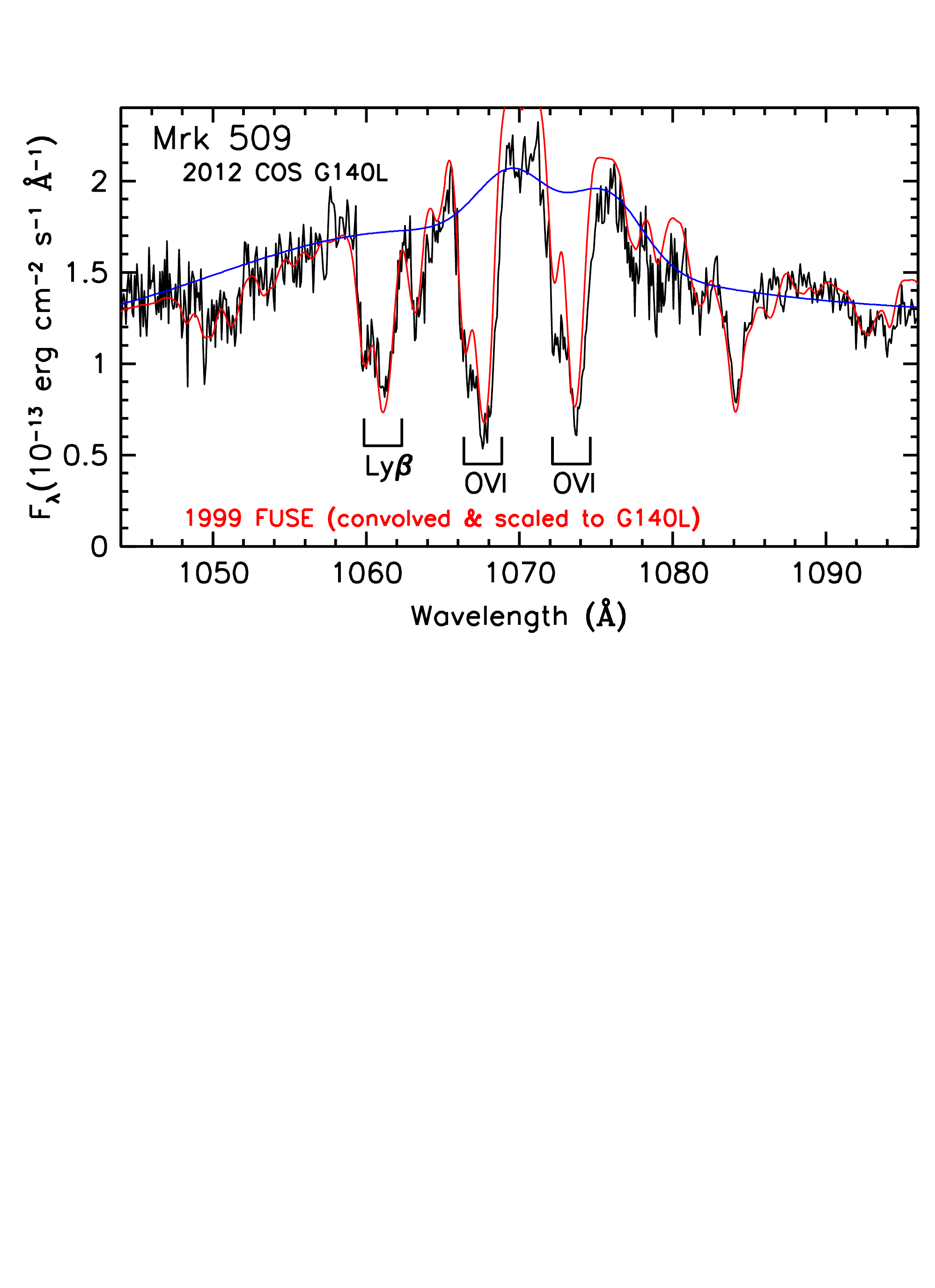}
  \vskip 65pt
  \caption{Top panel: Calibrated and merged COS G140L spectrum of Mrk~509 from 2012 (black) compared to the 1999 FUSE spectrum of \cite{Kriss00} (red).
The FUSE spectrum is convolved with the COS line-spread function to render it
at the same spectral resolution.
The best-fit emission model for the COS G140L is shown in blue.
Geocoronal emission at Ly$\beta$ in the FUSE spectrum
is indicated with an Earth symbol.
Bottom panel: The 2012 COS spectrum, its best-fit emission model,
and the 1999 FUSE spectrum are compared in greater detail in the region
surrounding the \ion{O}{vi} emission line.
The FUSE spectrum has been convolved to the COS resolution using the COS
line-spread function, and scaled up to match the flux in the red and blue wings
of the emission line. Note the higher peak in the emission line for the
FUSE spectrum and the subtle differences in the relative strengths of the
intrinsic absorption lines.}
  \label{fig:g140blue}
\end{figure*}

Below 1150 \AA, the blue-mode response of grating G140L permits a direct
comparison to the FUSE observations of the prior decade \citep{Kriss00}.
In 2012 Mrk 509 was significantly brighter than in 1999.
The top panel of Figure \ref{fig:g140blue}
compares the 2012 COS spectrum to the 1999 FUSE
spectrum of \cite{Kriss00}, with the FUSE data convolved to the COS
resolution using the LSF for G140L at 1068 \AA\ for
LP2 \citep{Duval13}.
These spectra are remarkably similar in shape,
consistent with the low degree of variability in spectral shape of Mrk 509,
although there are slight differences in the \ion{O}{vi} line profile and
absorption features as shown in
the bottom panel of Figure \ref{fig:g140blue}.
We will explore the variations in the absorption lines in more detail in
subsequent sections.

\begin{table*}
   \centering
	\caption[]{Emission Features in the 2012 COS Spectrum of Mrk~509}
	\label{tab:cos2012fit}
\begin{tabular}{l c c c c}
\hline\hline       
Feature & $\rm \lambda_0$ & Flux & $\rm v_{sys}$ & FWHM \\
  & ($\rm \AA$)  & ($\rm 10^{-14}~erg~cm^{-2}~s^{-1}~\AA^{-1}$) & ($\rm km~s^{-1}$) & ($\rm km~s^{-1}$) \\
\hline
\ion{C}{iii}  & 977.02 & $  6.1\pm 15.0$ & $ -10 \pm 340$ & $ 1530 \pm 260$\\
\ion{C}{iii}  & 977.02 & $ 25.0\pm 7.2$ & $ -40 \pm 210$ & $ 5580 \pm 690$\\
\ion{N}{iii}  & 989.80 & $ 33.0\pm 32.0$ & $  270 \pm 380$ & $ 8400 \pm 670$\\
Ly$\beta$   & 1025.72 & $ 170.0\pm 48.0$ & $  20 \pm 370$ & $ 8400 \pm 670$\\
\ion{O}{vi}  & 1031.93 & $ 23.0\pm 4.0$ & $  610 \pm 370$ & $ 1530 \pm 260$\\
\ion{O}{vi}  & 1037.62 & $ 23.0\pm 4.0$ & $  660 \pm 110$ & $ 1530 \pm 260$\\
\ion{O}{vi}  & 1031.93 & $ 30.0\pm 52.0$ & $  20 \pm 360$ & $ 8400 \pm 670$\\
\ion{O}{vi}  & 1037.62 & $ 15.0\pm 26.0$ & $  20 \pm 360$ & $ 8400 \pm 670$\\
Ly$\alpha$   & 1215.67 & $  8.9\pm 0.6$ & $  250 \pm 20$ & $ 650 \pm  60$\\
Ly$\alpha$   & 1215.67 & $ 75.0\pm 2.3$ & $ -170 \pm 20$ & $ 1330 \pm 250$\\
Ly$\alpha$   & 1215.67 & $ 380.0\pm 12.0$ & $ -240 \pm 20$ & $ 3090 \pm 630$\\
Ly$\alpha$   & 1215.67 & $ 820.0\pm 25.0$ & $ -230 \pm 20$ & $ 9860 \pm 650$\\
\ion{N}{v}   & 1238.82 & $ 25.0\pm 2.3$ & $  190 \pm 20$ & $ 1920 \pm 210$\\
\ion{N}{v}   & 1242.80 & $ 25.0\pm 2.3$ & $  180 \pm 20$ & $ 1920 \pm 210$\\
\ion{S}{ii}  & 1260.42 & $ 79.0\pm 2.5$ & $ -10 \pm 20$ & $ 6020 \pm 200$\\
\ion{O}{i}+\ion{S}{ii} & 1304.46 & $ 35.0\pm 1.2$ & $ -70 \pm 20$ & $ 3690 \pm 190$\\
\ion{C}{ii}  & 1334.53 & $ 10.0\pm 0.4$ & $  -10 \pm 30$ & $ 2320 \pm 230$\\
\ion{Si}{iv}  & 1393.76 & $ 79.0\pm 2.4$ & $  40 \pm 20$ & $ 4960 \pm 190$\\
\ion{Si}{iv}  & 1402.77 & $ 40.0\pm 1.2$ & $  40 \pm 20$ & $ 4960 \pm 190$\\
\ion{N}{iv}] & 1486.50 & $  3.6\pm 41.0$ & $ -20 \pm 50$ & $ 2180 \pm 210$\\
\ion{C}{iv}  & 1548.19 & $  5.2\pm 1.1$ & $  -80 \pm 40$ & $ 230 \pm  30$\\
\ion{C}{iv}  & 1550.77 & $  2.6\pm 0.5$ & $  -80 \pm 40$ & $ 230 \pm  30$\\
\ion{C}{iv}  & 1548.19 & $ 82.0\pm 4.1$ & $ -250 \pm 40$ & $ 2040 \pm 170$\\
\ion{C}{iv}  & 1550.77 & $ 41.0\pm 2.0$ & $ -250 \pm 40$ & $ 2040 \pm 170$\\
\ion{C}{iv}  & 1548.19 & $ 200.0\pm 8.5$ & $ -250 \pm 40$ & $ 4270 \pm 190$\\
\ion{C}{iv}  & 1550.77 & $ 99.0\pm 4.3$ & $ -250 \pm 40$ & $ 4270 \pm 190$\\
\ion{C}{iv}  & 1549.48 & $ 440.0\pm 25.0$ & $  -50 \pm 40$ & $11480 \pm 870$\\
\ion{He}{ii}  & 1640.45 & $  8.3\pm 58.0$ & $ -10 \pm 50$ & $ 1200 \pm 260$\\
\ion{He}{ii}  & 1640.45 & $ 71.0\pm 9.2$ & $ -30 \pm 170$ & $ 5040 \pm 260$\\
\hline
\end{tabular}
\end{table*}

\section{Data Analysis}

\subsection{Fitting the Continuum and Emission Lines}

To compare the intrinsic absorption lines in our 2012 spectra to the prior
epochs observed with FUSE and COS, we first fit an emission model to the
spectrum in order to produce normalized spectra.
As in \cite{Kriss11b}, we use a reddened power law to describe the continuum
shape and combinations of Gaussian emission components to model the
emission lines, with our fits optimized using the {\tt specfit} task
\citep{Kriss94} in IRAF.
As in our fits to the 2009 spectrum, we fix the extinction
at $\rm E(B-V)=0.057$ \citep{Schlafly11}
and the foreground damped Ly$\alpha$ absorption by the Milky Way at a
column density of $\rm N_H = 3.9 \times 10^{20}~cm^{-2}$ \citep{Wakker11}.
The best-fit power law is very similar to that of the COS 2009 spectrum, with
$\rm F_\lambda = 2.44 \times 10^{-13} (\lambda / 1000 \AA)^{-1.57})$
$\rm erg~cm^{-2}~s^{-1}~\AA^{-1}$.

As in our fits to the 2009 spectrum, we use four Gaussians to describe the
strongest lines (Ly$\alpha$, \ion{C}{iv}, and \ion{O}{vi})--a very broad base
with a full-width at half-maximum of $\sim$10,000 $\rm km~s^{-1}$,
two moderate width broad components with FWHM of a few thousand $\rm km~s^{-1}$,
and a narrow component with FWHM$\sim$300 $\rm km~s^{-1}$.
Weaker lines require only one or two Gaussians.
For lines comprised of doublets (\ion{O}{vi}, \ion{N}{v}, \ion{Si}{iv}, and
\ion{C}{iv}),
we include a component for each narrow-line component of the doublet with the
relative wavelengths fixed at the ratio of the laboratory values, and an
optically thin, 2:1 ratio for the blue to the red flux components.
Figure 4 in \cite{Kriss11b} illustrates a similar fit to the 2009 COS spectrum.
Table \ref{tab:cos2012fit} gives the components and best-fit parameters for
our model of the 2012 COS spectrum.

Many of the individual components of the spectral features in our fit
are significant, but their parameters are highly uncertain.
In the case of multi-component lines like \ion{C}{iii} $\lambda$977 this
is because the Gaussian decomposition we use is not unique.
If one takes the feature as a whole, its statistical significance is high,
as is the fact that it is not well described by a Gaussian with a single width.
In weak lines, one can easily trade flux between components and adjust their
widths and positions accordingly.
The quoted uncertainties reflect the formal errors associated with this process.
Since our overall goal is to determine a total emission model that enables us
to normalize the spectrum and measure the depths of the absorption lines,
these uncertainties are not an issue.

The best-fit emission model for the COS G140L spectrum is shown overlaying
the data in Figure \ref{fig:g140blue}. The model smoothly follows the
unabsorbed portions of the spectrum. In particular, note that in the regions
affected by intrinsic absorption in NGC 5548, the model is quite smooth.
Variations in the emission profile are on scales much larger than the
absorption troughs. Thus variations in the depths of the troughs that we will
discuss later are due to intrinsic variations in the absorption, not
variations in the emission model.

\subsection{Comparisons to Prior COS and FUSE Spectra}

\begin{figure}[!tbp]
\vspace{-0.25cm}
\centering
\hspace*{-0.25cm}\resizebox{1.05\hsize}{!}{\includegraphics[angle=0]{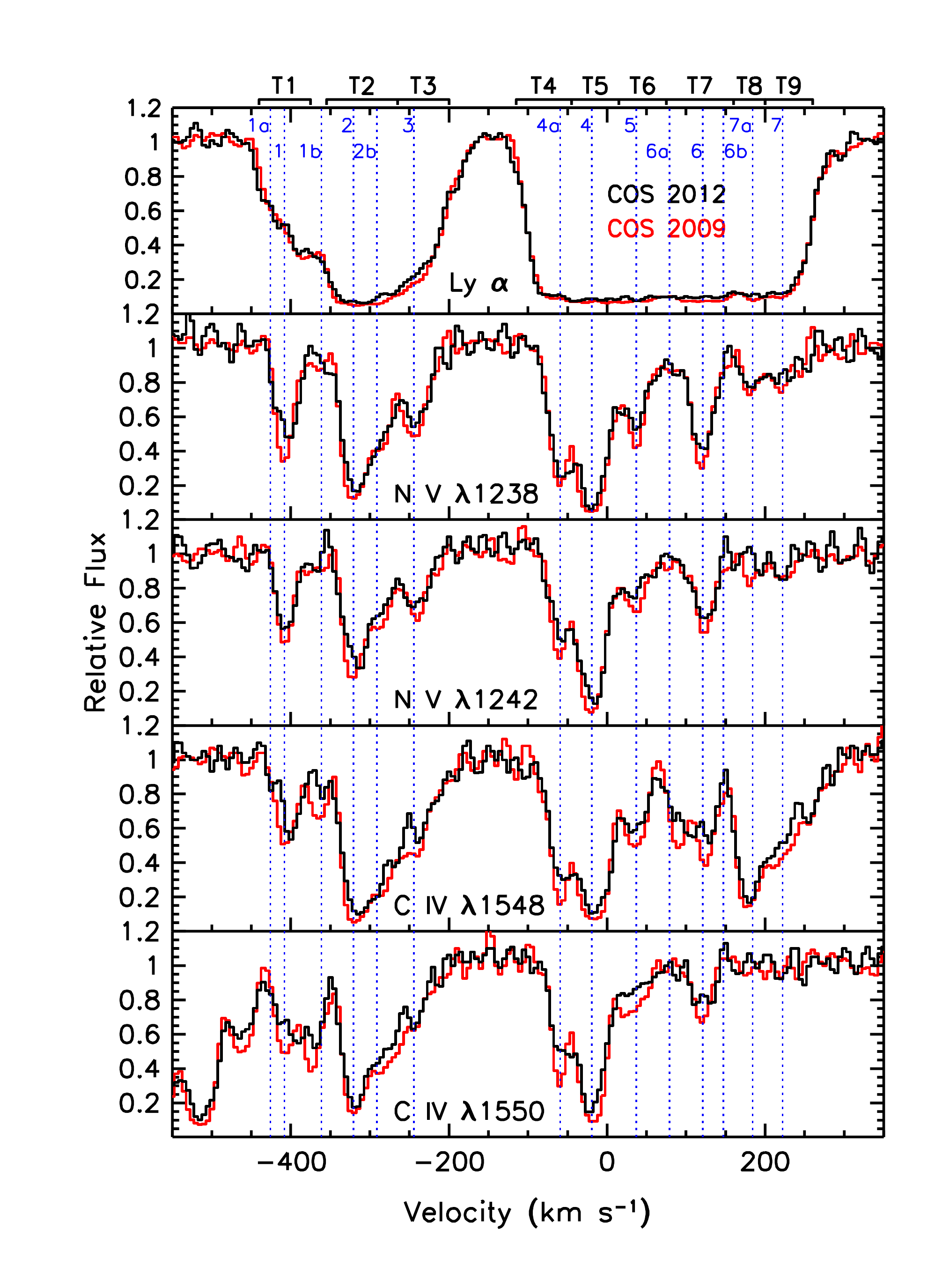}}
\caption{Normalized spectra comparing the intrinsic absorption features in
Mrk 509 from COS spectra obtained in 2009 (red) and 2012 (black).
Relative fluxes are plotted as a function of velocity relative to the
host galaxy redshift of $z = 0.034397$ \citep{Fisher95}.
The blue dotted vertical lines are labeled with the components identified by
\cite{Kriss11b}.
The boundaries of the selected spectral regions used to evaluate variability
in the absorption troughs are indicated along the top horizontal axis
(see also Table \ref{tbl_cos09cos12}).
}
\label{fig:2009vs2012}
\end{figure}

Using the emission model described in the last section, we constructed
normalized spectra by dividing the emission model
into the deconvolved 2012 COS spectra.
We can now directly compare these to the normalized deconvolved COS spectra
from 2009.
As in \cite{Arav12}, we resampled the G130M and G160M spectra onto a common
velocity scale using bins of $5~\rm km~s^{-1}$.
Figure \ref{fig:2009vs2012} compares the Ly$\alpha$, \ion{N}{v}, and
\ion{C}{iv} absorption troughs from these two epochs.

Overall, there are only minor changes in the depths of the
intrinsic absorption troughs.
To quantitatively evaluate the statistical significance of changes in the
absorption troughs, we followed the approach of \cite{Arav12} to compare the
average transmission of the selected spectral regions shown in
Figure \ref{fig:2009vs2012} (labeled T1--T9).
Table \ref{tbl_cos09cos12} compares the differences in transmission in each
of the absorption troughs T1--T9 between the two epochs.
In Ly$\alpha$, the most highly blueshifted troughs, T1 and T2, as well as
the main components in the red trough, T4 through T8, show statistically
significant variations.
The changes are in the sense of increased transmission (decreased absorption)
in 2012 compared to 2009.
Since \cite{Arav12} found that the optical depths of the \ion{C}{iv} and
\ion{N}{v} absorption lines in Mrk 509 were not saturated, these variations
can be attributed to changes in column density due to photoionization. 
However, given the high degree of saturation in the
Ly$\alpha$ absorption, it is not clear if these might be slight variations
in covering factor rather than an ionization response.
We therefore do not use Ly$\alpha$ in our analysis below.
In \ion{N}{v}, as in \cite{Arav12}, to consider a variation to be significant,
we require both the red and the blue components to show a consistent change
$> 2 \sigma$ between the COS observations in 2009 and 2012.
Under this criterion, we detect significant changes in troughs T1, T3, and
T5--T7 in  \ion{N}{v}. As in Ly$\alpha$, these are all increases in
transmission.

Evaluating \ion{C}{iv} is more difficult since the close pairing of the
doublet transitions ($500~\rm km~s^{-1}$) makes troughs T1--T3 of the red
doublet overlap with troughs T7--T9 of the blue doublet.
Similarly, trough T7 of the blue doublet is blended with trough T1
of the red doublet.
Therefore we can only apply the same criterion of confirmed variability in
both red and blue components to troughs T2--T6 for \ion{C}{iv};
all of these show increased transmission, as does the single uncontaminated
trough T1 in the blue line of the doublet.

\begin{figure}[!tbp]
\vspace{-0.25cm}
\centering
\hspace*{-0.25cm}\resizebox{1.05\hsize}{!}{\includegraphics[angle=0]{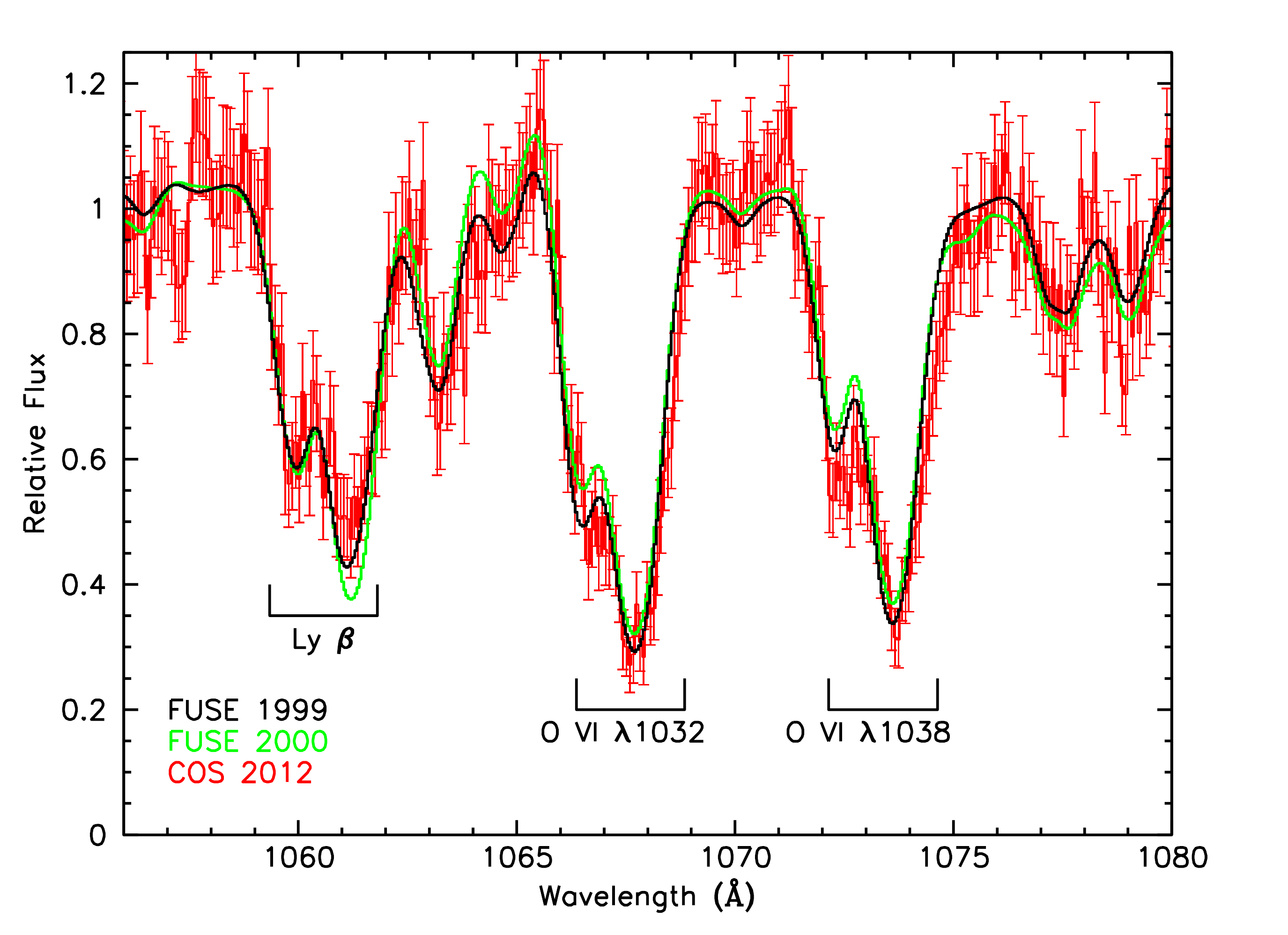}}
\caption{Normalized spectra comparing the intrinsic absorption features in
Mrk 509 from the COS G140L spectrum obtained in 2012 (red) to
FUSE spectra from 1999 (black) and 2000 (green).
1-$\sigma$ error bars are attached to each COS spectral point.
Relative fluxes are plotted as a function of observed wavelength.
The Ly$\beta$ and \ion{O}{vi} absorption troughs in Mrk 509 are labeled.
Other dips in the spectra are foreground galactic interstellar absorption lines.
}
\label{fig:cos2012vs_fuse_wave}
\end{figure}

\begin{figure}[!tbp]
\vspace{-0.25cm}
\centering
\hspace*{-0.25cm}\resizebox{1.05\hsize}{!}{\includegraphics[angle=0]{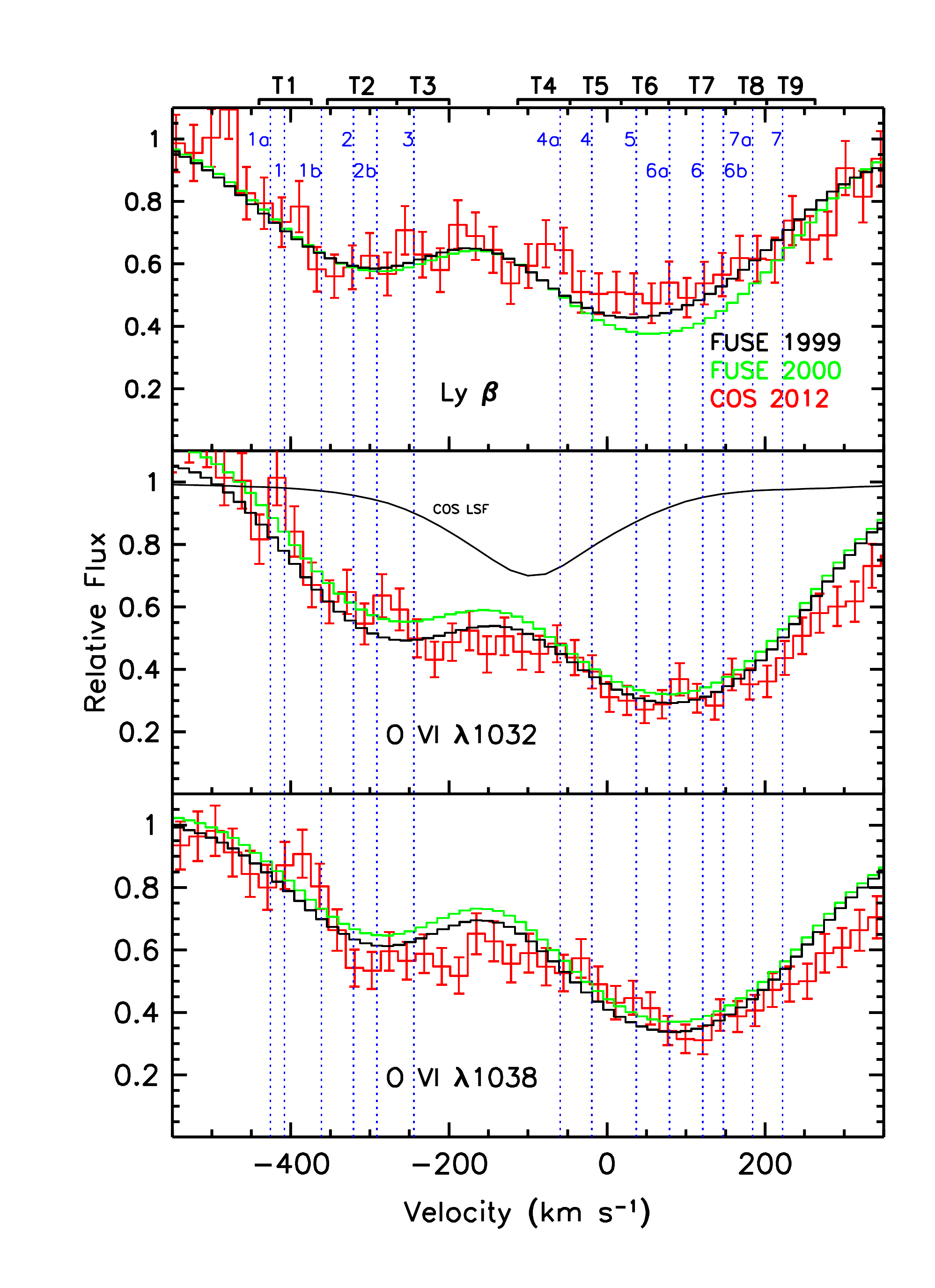}}
\caption{Normalized spectra comparing the intrinsic absorption features in
Mrk 509 from the COS G140L spectrum obtained in 2012 (red, with 1-$\sigma$
error bars) to
FUSE spectra from 1999 (black) and 2000 (green).
Relative fluxes are plotted as a function of velocity relative to the
host galaxy redshift of $z =0.034397$ \citep{Fisher95}.
The blue dotted vertical lines are labeled with the components identified by
\cite{Kriss11b} at higher spectral resolution.
As a guide to the expected widths of unresolved spectral features, the
COS line-spread function (LSF) is shown as the thin black line in the
middle panel.
The boundaries of the selected spectral regions used to evaluate variability
in the absorption troughs are indicated along the top horizontal axis
(see also Table \ref{tbl_cosfuse}).
}
\label{fig:cos2012vs_fuse_vel}
\end{figure}

These very small variations, although significant, are difficult to evaluate
in terms of photoionization models given our extremely sparse sampling.
Table \ref{tbl_fluxes} summarizes continuum flux measurements at a common
wavelength for the prior $\sim$15 years of spectroscopic observations of
Mrk 509.
Errors on the continuum fluxes in Table \ref{tbl_fluxes} are purely
statistical. Absolute calibration errors are on the order of 5\% for the COS
spectra, and 5--10\% for FUSE.
\cite{Arav12} found significant variations only in troughs T1 and T2
between the STIS 2001 observation and the COS 2009 observation.
While our higher signal-to-noise COS observations from 2012 allow us to detect
significant, but very slight changes from 2009, the continuum flux varied
hardly at all between the two observations.
Given the long response times inferred by  \cite{Arav12} for the outflow in
Mrk 509, these slight variations are more reflective of an integrated response
to variations over very long timescales, likely years.

\begin{table*}
  \centering
	\caption[]{Variations in Mrk 509 Absorption Troughs between 2009 and 2012}
	\label{tbl_cos09cos12}
\begin{tabular}{l c c c c c c c}
\hline\hline       
 & & & & & & & \\
Feature & Absorption & $v_1$$^{\mathrm{a}}$ & $v_2$$^{\mathrm{a}}$ & $<T_{COS12}>$$^{\mathrm{b}}$ & ${<T_{COS12} - T_{COS09}>}^{\mathrm{c}} \over {<T_{COS12}>}$ & ${\sigma(COS12-COS09)}^{\mathrm{d}} \over {<T_{COS12}>}$ & ${<T_{COS12} - T_{COS09}>}^{\mathrm{e}} \over  {\sigma(COS12-COS09)}$ \\
        & Trough     & ($\rm km~s^{-1}$) & ($\rm km~s^{-1}$)  & & &  & \\
\hline
Ly$\alpha$ & T1 & $-$440 & $-$375 & 0.504 & $-$0.001 & 0.0049 & $-$0.1 \\
Ly$\alpha$ & T2 & $-$355 & $-$265 & 0.106 &  0.212 & 0.0096 & 22.1 \\
Ly$\alpha$ & T3 & $-$265 & $-$200 & 0.323 &  0.114 & 0.0057 & 19.8 \\
Ly$\alpha$ & T4 & $-$115 &  $-$45 & 0.277 & $-$0.013 & 0.0064 & $-$2.1 \\
Ly$\alpha$ & T5 &  $-$45 &  $\phantom{0}$15 & 0.082 &  0.149 & 0.0126 & 11.8 \\
Ly$\alpha$ & T6 &  $\phantom{0}$15 &  $\phantom{0}$75 & 0.093 &  0.139 & 0.0110 & 12.6 \\
Ly$\alpha$ & T7 &  $\phantom{0}$75 &  160 & 0.101 &  0.175 & 0.0093 & 18.8 \\
Ly$\alpha$ & T8 &  160 &  200 & 0.113 &  0.113 & 0.0128 &  8.8 \\
Ly$\alpha$ & T9 &  200 &  260 & 0.208 &  0.079 & 0.0080 &  9.9 \\
\hline
\ion{N}{v} $\lambda1238$ & T1 & $-$440 & $-$375 & 0.767 &  0.031 & 0.0070 &  4.5 \\
\ion{N}{v} $\lambda1238$ & T2 & $-$355 & $-$265 & 0.470 &  0.004 & 0.0078 &  0.5 \\
\ion{N}{v} $\lambda1238$ & T3 & $-$265 & $-$200 & 0.741 &  0.017 & 0.0069 &  2.5 \\
\ion{N}{v} $\lambda1238$ & T4 & $-$115 &  $-$45 & 0.683 & $-$0.022 & 0.0070 & $-$3.1 \\
\ion{N}{v} $\lambda1238$ & T5 &  $-$45 &  $\phantom{0}$15 & 0.302 &  0.131 & 0.0121 & 10.9 \\
\ion{N}{v} $\lambda1238$ & T6 &  $\phantom{0}$15 &  $\phantom{0}$75 & 0.707 &  0.038 & 0.0073 &  5.2 \\
\ion{N}{v} $\lambda1238$ & T7 &  $\phantom{0}$75 &  160 & 0.737 &  0.032 & 0.0063 &  5.2 \\
\ion{N}{v} $\lambda1238$ & T8 &  160 &  200 & 0.832 & $-$0.008 & 0.0082 & $-$0.9 \\
\ion{N}{v} $\lambda1238$ & T9 &  200 &  260 & 0.864 & $-$0.010 & 0.0072 & $-$1.3 \\
\hline
\ion{N}{v} $\lambda1242$ & T1 & $-$440 & $-$375 & 0.819 &  0.015 & 0.0068 &  2.1 \\
\ion{N}{v} $\lambda1242$ & T2 & $-$355 & $-$265 & 0.663 &  0.069 & 0.0067 & 10.3 \\
\ion{N}{v} $\lambda1242$ & T3 & $-$265 & $-$200 & 0.846 &  0.040 & 0.0068 &  5.8 \\
\ion{N}{v} $\lambda1242$ & T4 & $-$115 &  $-$45 & 0.825 &  0.011 & 0.0068 &  1.6 \\
\ion{N}{v} $\lambda1242$ & T5 &  $-$45 &  $\phantom{0}$15 & 0.422 &  0.060 & 0.0108 &  5.5 \\
\ion{N}{v} $\lambda1242$ & T6 &  $\phantom{0}$15 &  $\phantom{0}$75 & 0.849 &  0.049 & 0.0071 &  6.9 \\
\ion{N}{v} $\lambda1242$ & T7 &  $\phantom{0}$75 &  160 & 0.855 &  0.043 & 0.0062 &  6.9 \\
\ion{N}{v} $\lambda1242$ & T8 &  160 &  200 & 0.956 &  0.053 & 0.0084 &  6.3 \\
\ion{N}{v} $\lambda1242$ & T9 &  200 &  260 & 0.967 &  0.040 & 0.0069 &  5.8 \\
\hline                  
\ion{C}{iv} $\lambda1548$ & T1 & $-$440 & $-$375 & 0.813 &  0.059 & 0.0059 & 10.0 \\
\ion{C}{iv} $\lambda1548$ & T2 & $-$355 & $-$265 & 0.365 &  0.099 & 0.0079 & 12.5 \\
\ion{C}{iv} $\lambda1548$ & T3 & $-$265 & $-$200 & 0.665 &  0.088 & 0.0065 & 13.6 \\
\ion{C}{iv} $\lambda1548$ & T4 & $-$115 &  $\phantom{0}-$45 & 0.682 &  0.030 & 0.0061 &  5.0 \\
\ion{C}{iv} $\lambda1548$ & T5 &  $\phantom{0}-$45 &  $\phantom{0}$15 & 0.284 &  0.058 & 0.0109 &  5.4 \\
\ion{C}{iv} $\lambda1548$ & T6 &  $\phantom{0}$15 &  $\phantom{0}$75 & 0.716 &  0.021 & 0.0063 &  3.3 \\
\ion{C}{iv} $\lambda1548$ & T7 &  $\phantom{0}$75 &  160 & 0.667 &  0.094 & 0.0057 & 16.6 \\
\ion{C}{iv} $\lambda1548$ & T8 &  160 &  200 & 0.329 &  0.106 & 0.0126 &  8.5 \\
\ion{C}{iv} $\lambda1548$ & T9 &  200 &  260 & 0.576 &  0.134 & 0.0075 & 17.9 \\
\hline                  
\ion{C}{iv} $\lambda1550$ & T1 & $-$440 & $-$375 & 0.697 &  0.056 & 0.0063 &  8.9 \\
\ion{C}{iv} $\lambda1550$ & T2 & $-$355 & $-$265 & 0.483 &  0.119 & 0.0067 & 17.7 \\
\ion{C}{iv} $\lambda1550$ & T3 & $-$265 & $-$200 & 0.813 &  0.081 & 0.0059 & 13.8 \\
\ion{C}{iv} $\lambda1550$ & T4 & $-$115 &  $\phantom{0}-$45 & 0.808 &  0.020 & 0.0058 &  3.5 \\
\ion{C}{iv} $\lambda1550$ & T5 &  $\phantom{0}-$45 &  $\phantom{0}$15 & 0.428 &  0.094 & 0.0092 & 10.3 \\
\ion{C}{iv} $\lambda1550$ & T6 &  $\phantom{0}$15 &  $\phantom{0}$75 & 0.894 &  0.067 & 0.0061 & 10.9 \\
\ion{C}{iv} $\lambda1550$ & T7 &  $\phantom{0}$75 &  160 & 0.928 &  0.010 & 0.0052 &  1.9 \\
\ion{C}{iv} $\lambda1550$ & T8 &  160 &  200 & 1.010 &  0.025 & 0.0071 &  3.6 \\
\ion{C}{iv} $\lambda1550$ & T9 &  200 &  260 & 0.998 &  0.008 & 0.0060 &  1.4 \\
\hline                  
\end{tabular}
\begin{list}{}{}
\item[{\bf Notes. }$^{\mathrm{a}}$]
Velocities are relative to a systemic redshift of $z = 0.034397$ \citep{Fisher95}.
\item[$^{\mathrm{b}}$]
Mean transmission in the given absorption trough in the COS 2012 spectrum.
\item[$^{\mathrm{c}}$]
Mean fractional difference between COS 2012 and COS 2009 troughs normalized
by the mean transmission for COS 2012.
\item[$^{\mathrm{d}}$]
Mean fractional error in the difference between the COS 2012 and COS 2009
troughs normalized by the mean transmission for COS 2012.
\item[$^{\mathrm{e}}$]
Mean fractional difference between the COS 2012  and COS 2009 troughs
normalized by the error.
\end{list}

\end{table*}

\begin{table}
  \centering
	\caption[]{Continuum Fluxes for Mrk 509}
	\label{tbl_fluxes}
\begin{tabular}{l c c}
\hline\hline       
 & &  \\
Observation & Date & $\rm F_{\lambda}(1160 \AA)$ \\
            &      & ($\rm 10^{-14}~erg~cm^{-2}~s^{-1}~\AA^{-1}$) \\
\hline
FUSE 1999 & 1999-11-04 & $7.7 \pm 0.7$ \\
FUSE 2000 & 2000-09-05 & $4.1 \pm 0.2$ \\
STIS 2000 & 2001-04-13 & $5.8 \pm 0.2$ \\
COS 2009 & 2009-12-10 & $8.8 \pm 0.2$ \\
COS 2012 & 2012-09-22 & $8.8 \pm 0.2$ \\
\hline                  
\end{tabular}
\end{table}

The variations we detected above are consistent with differences seen when
we now compare the
COS 2012 blue-mode spectra to prior FUSE observations from 1999 and 2000
\citep{Kriss00, Kriss11b}.
Figure \ref{fig:cos2012vs_fuse_wave}
compares the normalized COS 2012 G140L spectrum in the Ly$\beta$+\ion{O}{vi}
region to normalized FUSE spectra from the 1999 and 2000 observations,
which have been convolved to the resolution of the COS spectrum.

Significant differences between the COS and FUSE spectra are apparent in
the deepest, longer-wavelength portion of the Ly$\beta$ trough,
the shallower short wavelength portions of the \ion{O}{vi} troughs,
and the red wings of the \ion{O}{vi} troughs.
As with the \ion{N}{v} and \ion{C}{iv} troughs, we evaluate the statistical
significance of any variations by comparing the average transmissions as
integrated in the normalized spectra across the troughs labeled T1--T9 in
Figure \ref{fig:cos2012vs_fuse_vel}.
Table \ref{tbl_cosfuse} summarizes these differences.
Although we have used the same velocity bins for this analysis as we did for the
high resolution G130M and G160M spectra, we note that the
$\sim 150~\rm km~s^{-1}$ resolution of the blue-mode G140L grating really
does not resolve troughs T1--T3, or T4--T9.
Nevertheless, as shown in Table \ref{tbl_cosfuse}, the aggregate of troughs
T4--T8 comprising the heart of the redmost trough in the outflow in Ly$\beta$
all show significantly higher transmission in the COS 2012 spectrum compared to
the lower-flux FUSE 2000 observation.
In \ion{O}{vi}, the blue trough T3 shows significantly decreased
transmission in the COS 2012 spectrum compared to FUSE 2000, as do the red
troughs T4, T8 and T9.
Qualitatively, this is consistent in the sense of a photoionization response
since the COS 2012 continuum is more than twice as bright as that in the FUSE
2000 spectrum. The higher flux ionizes more of the neutral hydrogen,
increasing the transmission of the Ly$\beta$ trough.
This higher ionization state is also consistent with a rise in the ion
fraction of \ion{O}{vi} for the photoionization models of
\cite{Kraemer03} and \cite{Arav12}.

\begin{table*}
  \centering
	\caption[]{Variations in Mrk 509 Absorption Troughs between 2000 and 2012}
	\label{tbl_cosfuse}
\begin{tabular}{l c c c c c c c}
\hline\hline       
 & & & & & & & \\
Feature & Absorption & $v_1$$^{\mathrm{a}}$ & $v_2$$^{\mathrm{a}}$ & $<T_{COS12}>$$^{\mathrm{b}}$ & ${<T_{COS12} - T_{FUSE00}>}^{\mathrm{c}} \over {<T_{COS12}>}$ & ${\sigma(COS12-FUSE00)}^{\mathrm{d}} \over {<T_{COS12}>}$  & ${<T_{COS12} - T_{FUSE00}>}^{\mathrm{e}} \over  {\sigma(COS12-FUSE00)}$ \\
        & Trough     & ($\rm km~s^{-1}$) & ($\rm km~s^{-1}$)  & & &  & \\
\hline
Ly$\beta$ & T1 & $-$440 & $-$374 & 0.742 &  0.033 & 0.041 &  0.8 \\
Ly$\beta$ & T2 & $-$354 & $-$266 & 0.599 &  0.011 & 0.038 &  0.3 \\
Ly$\beta$ & T3 & $-$266 & $-$200 & 0.624 &  0.032 & 0.046 &  0.7 \\
Ly$\beta$ & T4 & $-$113 &  $\phantom{0}-$47 & 0.621 &  0.126 & 0.047 &  2.7 \\
Ly$\beta$ & T5 &  $\phantom{0}-$48 &  $\phantom{0}$18 & 0.532 &  0.186 & 0.053 &  3.5 \\
Ly$\beta$ & T6 &  $\phantom{0}$12 &  $\phantom{0}$78 & 0.507 &  0.243 & 0.064 &  3.8 \\
Ly$\beta$ & T7 &  $\phantom{0}$74 &  162 & 0.544 &  0.226 & 0.045 &  5.0 \\
Ly$\beta$ & T8 &  158 &  202 & 0.610 &  0.146 & 0.049 &  3.0 \\
Ly$\beta$ & T9 &  197 &  263 & 0.666 &  0.018 & 0.047 &  0.4 \\
\hline                  
\ion{O}{vi} $\lambda1032$ & T1 & $-$440 & $-$374 & 0.838 &  0.003 & 0.047 &  0.1 \\
\ion{O}{vi} $\lambda1032$ & T2 & $-$354 & $-$266 & 0.610 &  0.009 & 0.044 &  0.2 \\
\ion{O}{vi} $\lambda1032$ & T3 & $-$266 & $-$200 & 0.509 & $-$0.104 & 0.052 & $-$2.0 \\
\ion{O}{vi} $\lambda1032$ & T4 & $-$113 &  $-$47 & 0.454 & $-$0.123 & 0.048 & $-$2.6 \\
\ion{O}{vi} $\lambda1032$ & T5 &  $\phantom{0}-$48 &  $\phantom{0}$18 & 0.375 & $-$0.075 & 0.052 & $-$1.4 \\
\ion{O}{vi} $\lambda1032$ & T6 &  $\phantom{0}$12 &  $\phantom{0}$78 & 0.291 & $-$0.157 & 0.064 & $-$2.5 \\
\ion{O}{vi} $\lambda1032$ & T7 &  $\phantom{0}$74 &  162 & 0.330 & $-$0.050 & 0.058 & $-$0.9 \\
\ion{O}{vi} $\lambda1032$ & T8 &  158 &  202 & 0.366 & $-$0.170 & 0.079 & $-$2.2 \\
\ion{O}{vi} $\lambda1032$ & T9 &  197 &  263 & 0.457 & $-$0.209 & 0.049 & $-$4.3 \\
\hline                  
\ion{O}{vi} $\lambda1038$ & T1 & $-$440 & $-$374 & 0.851 &  0.022 & 0.031 &  0.7 \\
\ion{O}{vi} $\lambda1038$ & T2 & $-$354 & $-$266 & 0.610 & $-$0.101 & 0.036 & $-$2.8 \\
\ion{O}{vi} $\lambda1038$ & T3 & $-$266 & $-$200 & 0.569 & $-$0.192 & 0.043 & $-$4.5 \\
\ion{O}{vi} $\lambda1038$ & T4 & $-$113 &  $\phantom{0}-$47 & 0.552 & $-$0.132 & 0.044 & $-$3.0 \\
\ion{O}{vi} $\lambda1038$ & T5 &  $\phantom{0}-$48 &  $\phantom{0}$18 & 0.498 &  0.033 & 0.047 &  0.7 \\
\ion{O}{vi} $\lambda1038$ & T6 &  $\phantom{0}$12 &  $\phantom{0}$78 & 0.407 &  0.034 & 0.061 &  0.5 \\
\ion{O}{vi} $\lambda1038$ & T7 &  $\phantom{0}$74 &  162 & 0.350 & $-$0.119 & 0.054 & $-$2.2 \\
\ion{O}{vi} $\lambda1038$ & T8 &  158 &  202 & 0.414 & $-$0.129 & 0.054 & $-$2.4 \\
\ion{O}{vi} $\lambda1038$ & T9 &  197 &  263 & 0.478 & $-$0.211 & 0.046 & $-$4.5 \\
\hline                  
\end{tabular}
\begin{list}{}{}
\item[{\bf Notes. }$^{\mathrm{a}}$]
Velocities are relative to a systemic redshift of $z = 0.034397$ \citep{Fisher95}.
\item[$^{\mathrm{b}}$]
Mean transmission in the given absorption trough in the COS 2012 spectrum.
\item[$^{\mathrm{c}}$]
Mean fractional difference between COS 2012 and FUSE 2000 troughs normalized
by the mean transmission for COS 2012.
\item[$^{\mathrm{d}}$]
Mean fractional error in the difference between the COS 2012 and FUSE 2000
troughs normalized by the mean transmission for COS 2012.
\item[$^{\mathrm{e}}$]
Mean fractional difference between the COS 2012 and FUSE 2000 troughs
normalized by the error.
\end{list}

\end{table*}

Since we detect variations in the transmission of all absorption troughs
in Mrk~509 in our most recent COS observation,
we can use the timescale of these variations to set lower limits
on the density of each of the absorption components based on the
recombination timescale \citep{KK95, Nicastro99, Arav12}.
Given the slight variations we observe in column density, we use the same
baseline photoionization solutions for each absorption trough as \cite{Arav12},
and use the associated timescales per electron number density
for these solutions for \ion{C}{iv} and \ion{N}{v} as given in their Table 6.
For troughs 8 and 9 in \ion{O}{vi},
we use the photoionization solution of \cite{Kraemer02},
and obtain the timescale per electron number density for
\ion{O}{vi} from the {\sc Cloudy} output \citep{Ferland17}.

For our sparsely sampled data, the fractional variation in the
continuum between our two widely spaced observations in 2009 and 2012 is
unknown. The actually observed variation is only 0.6\%,
but it is possible that Mrk 509 dropped in flux by 100\%
immediately after the COS 2009 observation, and then recovered to normal
fluxes by 2012. We therefore use this conservative assumption.
Since the gas density must be high enough to respond to such variations at
least on the timescale corresponding to the interval between our two
observations, we can set a lower limit on the density, which then
translates into an upper limit on the distance.

In Table \ref{tbl_distance} we give upper limits on the density as
determined by \cite{Arav12} from the lack of variations observed between
2001 and 2009 for both \ion{C}{iv} and \ion{N}{v}, and lower limits on the
density as determined from the variations we observed between 2009 and 2012.
These both translate into lower and upper limits on the distance of the
absorbing gas from the nucleus.
For the lower limits, we use the 99\% confidence limits given in Table 8
of \cite{Arav12}.
For the upper limits, we use the most stringent (smallest) distance implied
by our lower limits on the density determined by either the \ion{C}{iv},
\ion{N}{v}, or \ion{O}{vi} measurements.
Although the bounds on radial distance given in the last two columns of
Table \ref{tbl_distance} appear to locate the absorbing gas quite precisely,
our sparse sampling in time and many assumptions suggest great caution in
jumping to this conclusion.
In particular,
we note that for some components, e.g., T3, T6, and T7, our upper and lower
bounds on density and distance are incommensurate.
We attribute this to the statistical nature of our analysis and the sparse
sampling. The lower limits on distance derived from the Monte Carlo
simulations in \cite{Arav12} are based mostly on the lack of variations
detected between the STIS observations and the COS observations.
Not observing variations is a function of both the actual strength of the
change in absorption as well as having observations sufficiently sensitive to
detect variations. The STIS 2001 observation has significantly less
signal-to-noise than either COS observation; more sensitive observations in
2001 may well have led to detection of variations at the level we observed
between 2009 and 2012.
Given these uncertainties, it is quite likely that the lower limits on distance
can easily be much lower than the 99\% confidence limits of \cite{Arav12}.

\begin{table*}
\centering
\caption{Limits on Density and Distance for Mrk 509 Absorption Troughs}
	\label{tbl_distance}
\begin{tabular}{l c c c c c c c c c}
\hline\hline
 &  & 
\multicolumn{2}{c}{\ion{C}{iv}} &
\multicolumn{2}{c}{\ion{N}{v}} &
\multicolumn{2}{c}{\ion{O}{vi}} &   &   \\
\hline
Trough & Velocity & 
$n_{e,low}^{\rm a}$ & $n_{e,high}$ & 
$n_{e,low}^{\rm a}$ & $n_{e,high}$ & 
$n_{e,low}$ & $n_{e,high}$ & 
$R_{low}^{\rm b}$ & $R_{high}^{\rm c}$ \\
 & ($\rm km~s^{-1}$) &
   ($\rm log~cm^{-3}$) & ($\rm log~cm^{-3}$) & 
   ($\rm log~cm^{-3}$) & ($\rm log~cm^{-3}$) & 
   ($\rm log~cm^{-3}$) & ($\rm log~cm^{-3}$) & 
   (pc) & (pc) \\
(1) & (2) & (3) & (4) & (5) & (6) & (7) & (8) & (9) & (10)\\
\hline
T1 & $-408$ & $\gtsim$3.3 & \ldots & $\gtsim$2.8 & $\ltsim$4.2 & \ldots & \ldots & $\gtsim \phantom{0}$60 & $\ltsim$130 \\
T2 & $-310$ & $\gtsim$3.0 & $\ltsim$3.2 & \ldots & $\ltsim$4.2 & \ldots & \ldots & $\gtsim$160 & $\ltsim$170 \\
T3 & $-233$ & $\gtsim$2.6 & $\ltsim$3.2 & $\gtsim$2.6& $\ltsim$3.9 & $\gtsim$1.6 & \ldots & $\gtsim$130 & $\ltsim$130 \\
T4 & $-80$ & $\gtsim$2.5 & $\ltsim$3.1 & \ldots & $\ltsim$3.4 & $\gtsim$1.6 & \ldots & $\gtsim$130 & $\ltsim$220 \\
T5 & $-15$ & $\gtsim$2.5 & $\ltsim$3.6 & $\gtsim$2.8 & \ldots & \ldots & \ldots & $\gtsim$130 & $\ltsim$220 \\
T6 & $+45$ & $\gtsim$2.3 & $\ltsim$2.8 & $\gtsim$3.1 & $\ltsim$3.1 & \ldots & \ldots & $\gtsim$150 & $\ltsim \phantom{0}$90 \\
T7 & $+118$ & \ldots & $\ltsim$2.5 & $\gtsim$2,4 & \ldots & \ldots & \ldots & $\gtsim$130 & $\ltsim$120 \\
T8 & $+180$ & $\gtsim$1.8 & \ldots & \ldots & \ldots & $\gtsim$2.3 & \ldots & \ldots & $\ltsim$140 \\
T9 & $+230$ & \ldots & \ldots & \ldots & \ldots & $\gtsim$2.3 & \ldots & \ldots & $\ltsim$140 \\
\hline
\end{tabular}
\begin{list}{}{}
\item[{\bf Notes. }]
Column (1): Absorption trough as defined in Table 4.
Column (2): Central velocity of the trough ($\rm km~s^{-1})$.
Column (3): Lower limit on the density derived from \ion{C}{iv} ($\rm log~cm^{-3}$).
Column (4): Upper limit on the density derived from \ion{C}{iv} ($\rm log~cm^{-3}$).
Column (5): Lower limit on the density derived from \ion{N}{v} ($\rm log~cm^{-3}$).
Column (6): Upper limit on the density derived from \ion{N}{v} ($\rm log~cm^{-3}$).
Column (7): Lower limit on the density derived from \ion{O}{vi} ($\rm log~cm^{-3}$).
Column (8): Upper limit on the density derived from \ion{O}{vi} ($\rm log~cm^{-3}$).
Column (9): Lower limit on the distance (pc).
Column (10): Upper limit on the distance (pc).\\
$^{\rm a}$\,Upper limits on density taken from Table 8 of \cite{Arav12}.\\
$^{\rm b}$\,Lower limits on distance taken from the 99\% confidence limits in
Table 8 of \cite{Arav12}.\\
$^{\rm c}$\,Upper limits on distance use the most stringent (highest) limits on
the density from either \ion{C}{iv}, \ion{N}{v}, or \ion{O}{vi}.
\end{list}

\end{table*}

\subsection{No Features Related to Ultra-fast Outflows}

Mrk 509 is among the growing number of AGN in which ultra-fast outflows (UFO)
have been detected in X-ray observations \citep{Tombesi10, Gofford13}.
These outflows typically have velocities exceeding 10,000 $\rm km~s^{-1}$ and
high column densities $(> 10^{23}~\rm cm^{-2})$.
Due to their high ionization,
they often only show absorption in \ion{Fe}{xxv} or \ion{Fe}{xxvi}.
Despite their high ionization, it is possible for X-ray UFOs to show related
absorption in \ion{H}{i} Ly$\alpha$ due to the high total column density of
the UFO, as seen in the quasar PG1211+143 \citep{Danehkar18}.
As with the X-ray absorption features, these UV counterparts are also
variable \citep{Kriss18a}.

In Mrk 509, \cite{Tombesi10} reported three systems with
$v_{out} = -0.141 c, -0.172 c, \mathrm{and} -0.196 c$.
\cite{Ponti09} reported yet a fourth system at $v_{out} = -0.141 c$ in the
stacked {\it XMM-Newton} spectra.
In the 2009 {\it XMM-Newton} campaign on Mrk 509, no UFO features were seen
\citep{Ponti13}. 
Similarly, in the {\it Chandra} observations coordinated with these HST
observations, UFOs were also not detected.
\cite{Kriss18b} searched for Ly$\alpha$ counterparts to these X-ray UFOs in the
archival FUSE observations of Mrk 509 and found none.

Given the strong variability of UFOs, we have searched our new
HST spectra for possible counterparts. The lower resolution of the COS G140L
grating compared to FUSE makes this more difficult due to the unresolved
absorption from many interstellar absorption lines. However, when we compare
our G140L spectrum to the FUSE spectra convolved with the G140L
line-spread function and scaled to the G140L flux levels,
we find no differences.
We therefore set upper limits for Ly$\alpha$ counterparts to the
previously detected X-ray UFOs at the same levels as reported in
\cite{Kriss18b}.
For the three features at
$v_{out} = -0.141 c, -0.172 c, \mathrm{and} -0.196 c$,
we set $2 \sigma$ confidence level (95\%) upper limits on the \ion{H}{i}
column density of $\mathrm{N_H} < 1.1 \times 10^{13}~\mathrm{cm}^{-2},
7.2 \times 10^{12}~\mathrm{cm}^{-2},~\mathrm{and}~ 1.3 \times 10^{14}~\mathrm{cm}^{-2}$, respectively.
The feature at $v_{out} = -0.048 c$ falls at 1197 \AA, within the bandpass of
our high-resolution, high S/N G130M spectra. No absorption is found
beyond the usual narrow interstellar features of
\ion{Mn}{ii} $\lambda\lambda1197,1199$ and the \ion{N}{i} triplet at 1200 \AA.
For a broad absorption feature with FWHM=1000 $\rm km~s^{-1}$, we set an upper
limit on the equivalent width of 0.05 \AA, and an upper limit on the
\ion{H}{i} column density of $9.2 \times 10^{12}~\mathrm{cm}^{-2}$.

\section{Discussion}

Our HST/COS observations of Mrk 509 in 2012 supplement the
extensive multiwavelength campaign carried out with XMM-{\it Newton}, 
{\it Chandra}, and HST in 2009 \citep{Kaastra11a}.
These new FUV spectra sample variations in the intrinsic UV absorption troughs
on both shorter timescales ($\sim$3 years) and longer ones ($\sim13$ years)
than our original campaign.
The shorter timescale variations permit us to put better upper limits on the
location of the outflowing gas than was possible in \cite{Kriss11b}
and \cite{Arav12}.

\cite{Kriss11b} detected variations in absorption Component \#6 of Ly$\beta$,
which allowed them to set an upper limit of 1.25 kpc on its distance.
This absorption component comprises the bulk of trough T7 in \cite{Arav12}.
The lack of variability in this trough in both \ion{N}{v} and \ion{C}{iv}
between the STIS 2001 observation and the COS 2009 spectrum led to a lower
limit on its distance of $> 130$ pc at the 99\% confidence level,
consistent with the 1.25 kpc upper limit from Ly$\beta$ variability.
However, our new COS observations in 2012 do detect variability in trough T7
on shorter timescales than those sampled between the STIS 2001 and COS 2009
observations. This illustrates the difficulty of drawing firm conclusions based
on poorly sampled data. Lack of variability does not always imply a firm
upper limit on the gas density. It can be simply by chance that sparse
observations observe the same transmission, even though the absorption troughs
are varying on more rapid timescales.
In the case of trough T7, the change we see in the 3 years between the two
COS observations imply a minimum density of log~$n_e ~\rm cm^{-3} > 2.4$ and a
corresponding upper limit on its distance of 120 pc, comparable to the
upper limit based on non-detection.
Our new COS G140L blue-mode observations of the Ly$\beta$+\ion{O}{vi} region
confirm variability at these velocities in trough T7 (which includes
Component \#6), but on the longer timescale of 12 years, compared to 2 years.
This Ly$\beta$ variability does not improve our upper limit on the
distance, but it does confirm the detection of the column density variation
we see in \ion{N}{v}.

The outflow as manifested in the UV absorption lines is centrally concentrated
in the near nuclear region.
All absorption troughs have distances within $<$220 pc of the nucleus.
At the redshift of Mrk~509 ($z = 0.03937$), a Hubble constant of
$H_o = 70~\rm km~s^-1~Mpc^{-1}$, $\Omega_{DM} = 0.3$,
and $\Omega_\Lambda = 0.7$,
in the rest frame of the cosmic microwave background, the angular scale at
the distance of Mrk~509 is 667 $\rm pc~arcsec^{-1}$.
This places the troughs at angular scales of $<0.3''$, which corresponds to
the unresolved, highest surface brightness central peak in the optical and
near-IR IFU observations of Mrk~509 \citep{Fischer15, Liu15}.
Given that this region is unresolved, the UV absorption troughs could 
correspond to filaments observed at similar distances in nearby AGN such as
NGC 4151 \citep{Das05} or NGC 1068 \citep{Das07}.
As suggested by \cite{Crenshaw12}, the kinematics, ionization state, and
densities of these high-excitation emission-line knots and filaments are
very similar to the physical characteristics of the narrow UV absorption lines
observed in most low-redshift AGN.

The high blue-shifted velocities of troughs T1--T3
do not correspond to typical velocities observed in the IFU images, which
range $\sim -250$--$+250~\rm km~s^{-1}$ \citep{Fischer15, Liu15}, nor do they
lie at the highest distances (up to $\sim$1.5 kpc) seen in the IFU images.
The lower velocities at larger distances observed in the IFU images resemble
the behavior seen in the resolved structures near the centers of
NGC 4151 and NGC 1068.
The inner clouds have increasing velocity with distance until reaching a
maximum at $\sim$100 pc. At larger distances, velocities decrease.
This might suggest that the outflowing gas becomes mass loaded as it entrains
surrounding material from the ambient ISM, and then decelerates
\citep{Das05, Das07}.
If similar mechanisms are at work in Mrk 509, the overall impact on the host
galaxy is small.
\cite{Liu15} note that a small region northeast of the nucleus in Mrk~509
appears to have suppressed star formation, perhaps due to the impact of the
AGN outflow. However, there are many other star-forming regions at larger
radii that appear normal. So, the visible impact of the AGN outflow on the
overall character of the host galaxy appears to be minimal.

\section{Conclusions}

Sensitive HST/COS observations of Mrk 509 in 2012 following the 
multiwavelength campaign of \cite{Kaastra11a} allow us to detect variations
in absorption in the AGN outflow when compared to prior UV spectral
observations.
We detect significant changes on a shorter 3-year timescale that allow us to
set upper limits on the distance of most absorption troughs at distances of
100--200 pc.
This is consistent with the location established for the X-ray warm absorbers
(5--400 pc) by \cite{Kaastra12}.
It also corresponds to the unresolved central peak in IFU observations of the
narrow-line region of Mrk 509.
The nuclear outflow appears to have most of its impact confined to the
near-nuclear region, and have little impact on the host galaxy overall.

\begin{acknowledgements}
Based on observations made with
the NASA/ESA {\it HST}, and obtained from the Hubble
Legacy Archive.
This work was supported by NASA through grants for HST program number 12916
from the Space Telescope Science Institute, which is operated
by the Association of Universities for Research in Astronomy, Incorporated,
under NASA contract NAS5-26555.
SRON is supported financially by NWO, the Netherlands Organization for Scientific Research.
SB acknowledges financial support from the Italian Space Agency under grant
ASI-INAF I/037/12/0, and n. 2017-14-H.O.
EC is partially supported by the NWO-Vidi grant number 633.042.525.
GP acknowledges financial support from the Bundesministerium 
f\"{u}r Wirtschaft und Technologie/Deutsches Zentrum f\"{u}r Luft- 
und Raumfahrt (BMWI/DLR, FKZ 50 OR 1715 and FKZ 50 OR 1812) 
and the Max Planck Society.
POP acknowledges financial support from the High Energy National Program (PNHE) of CNRS and from the CNES agency.
\end{acknowledgements}

\bibliographystyle{aa}
\bibliography{mrk509}

\end{document}